\documentclass[fleqn]{2020SCGE}
 \usepackage {color}
\usepackage{ulem}

\begin{document}

\ensubject{subject}

\ArticleType{Article}

\title{First extraction of the proton mass radius and scattering length $\left|\alpha_{\rho^0 p}\right| $ from $\rho^0$ photoproduction}{}

\author[1,2]{Xiao-Yun Wang}{{xywang@lut.edu.cn}}%
\author[1,3,4]{Fancong Zeng}{{zengfc@ihep.ac.cn}}
\author[1]{Quanjin Wang}{}
\author[1]{Li Zhang}{}

\AuthorMark{Xiao-Yun Wang}

\AuthorCitation{X.Y. Wang, F. Zeng, Q. Wang, L. Zhang}

\address[1]{Department of physics, Lanzhou University of Technology,
Lanzhou 730050, China}
\address[2]{Lanzhou Center for Theoretical Physics, Key Laboratory of Theoretical Physics of Gansu Province, Lanzhou University, Lanzhou, Gansu 730000, China}

\address[3]{Institute of High Energy Physics, Chinese Academy of Sciences, Beijing 100049, China}
 
\address[4]{University of Chinese Academy of Sciences, Beijing 100049, China}


\abstract{
As it involves the lightest physical states excited from the vacuum by the vector quark current, near-threshold $\rho^0$ photoproduction is considered a possible 
way to research the proton radius and the absolute value of the scattering lengths of the $\rho^0$--proton interaction. In this work, under the assumption of a scalar form factor of dipole form, the value of the proton mass radius is calculated as $0.85\pm 0.06 \text{ fm }$ by fitting the differential cross section of the $\gamma p \rightarrow \rho^0 p$ reaction at near-threshold energy. For light vector meson photoproduction, because the exchange of a scalar quark--antiquark pair is not suppressed and should dominate the scalar gluon exchange, the radius we extract from $\rho^0$ photoproduction is likely to represent the quark radius of the proton. This fact may explain why the value obtained in this work is very near the proton charge radius. Moreover, the absolute value of the $\rho^0$--proton scattering length $|\alpha_{\rho^0 p}|= 0.31 \pm 0.06 \text{ fm}$ is obtained for the first time within the vector meson dominance model. This result disobeys the rule that the absolute value of the vector meson and proton scattering length $|\alpha_{V p}|$ increases with the meson's mass, which can be attributed to treating the $\rho^0$ meson as a point in the analysis. These results provide useful theoretical information for an in-depth understanding of proton structure and the proton--vector meson interaction.

 }%



\maketitle


\begin{multicols}{2}
 
\section{Introduction}
A precise determination of $\rho^0 (770)$ photoproduction at near-threshold energies has been motivated by several aspects. First, as mentioned in Ref. \cite{Kharzeev:2021qkd}, because the $\rho^0$ meson is the lightest physical state excited from the vacuum by the vector quark current, its near-threshold photoproduction probably contains a wealth of physical information. Furthermore, because it involves the same vector quantum numbers as the photon, namely, $J^{P C}=1^{--}$, the behavior of the near-threshold cross section can be related to the vector meson--proton ($V p$) scattering length \cite{Gell-Mann:1961jim}. It should be possible to extract the $\rho^0$-hadron scattering length from the near-threshold cross section of $\rho^0$ photoproduction. Moreover, theoretical studies indicate that near-threshold vector meson photoproduction will give access to many interesting physics aspects, such as a trace anomaly contribution to the nucleon mass and cusp effects \cite{Lutz:2001mi,Titov:2007xb}.

The proton radius is a large inspiration in understanding proton structure, and it can be measured by using an electron or photon as a probe. The most studied proton charge radius was obtained by colliding the nucleus with a high-energy electron.
The electromagnetic form factors provide necessary information about the distribution of electric charge \cite{Kohl:2008zz,Perdrisat:2006hj}. On the experimental side, the proton charge radius was determined as $0.8409 \pm 0.0004 $ fm \cite{Workman:2022ynf} from elastic electron scattering experiments. In addition to the charge radius extracted from the electromagnetic form factors, the mass radius reflects the mass distribution of the proton, which is a fundamental property of the proton \cite{Kharzeev:2021qkd}. However, little information about the proton mass radius is explicit at present. Because the interaction of gravitons and proton scattering is very weak, it is far beyond the measurement limit of current experiments. 
Fortunately, gravitational form factors (GFFs) are helpful for understanding the perturbative and nonperturbative quantum chromodynamics (QCD) effects, providing a connection to the spatial distribution of quarks inside the proton \cite{Kobzarev:1962wt,Pagels:1966zza,Ji:1996nm}.
In QCD theory, the photoproduction of a quarkonium off a proton is connected to the scalar GFFs, which are sensitive to the proton mass distribution from the QCD trace anomaly \cite{Wang:2019mza,Ji:1996nm}. Under an assumption of a scalar form factor of dipole form, the proton mass radius can be extracted through the near-threshold photoproduction of a vector meson differential cross section \cite{Kharzeev:2021qkd,Fujii:1999xn}.
Moreover, our previous work \cite{Wang:2022vhr} shows that extracting the proton mass radius from the near-threshold differential cross section of heavy quarkoniums is always affected by a large $|t|_{\min}$. Additionally, if the proton mass radius is extracted from the lighter vector meson photoproduction data, the effect from $|t|_{\min}$ will be much smaller. Thus, as it involves a much lighter meson, the $\rho^0$ photoproduction process may be worth choosing to calculate the proton mass radius from this perspective.

The evaluation of the scattering lengths may serve as a unique input for QCD-motivated models of vector meson--nucleon interactions \cite{Strakovsky:2014wja,Strakovsky:2020uqs,Strakovsky:2019bev,Strakovsky:2021vyk}.
The behavior of the near-threshold cross section is related to the $V p $ scattering length \cite{Gell-Mann:1961jim}. Provided in the vector meson dominance (VMD) model \cite{ Titov:2007xb,Kroll:1967it}, the absolute value of the scattering lengths for $\omega p, \phi p, J/\psi p, \psi(2S) p $, and $ \Upsilon p$ reactions have been reported by using the recent photoproduction experiment data or quasi data \cite{Strakovsky:2014wja,Strakovsky:2020uqs,Strakovsky:2019bev,Strakovsky:2021vyk,Wang:2022xpw}. Because the VMD model does not contain free parameters in the process from the $\gamma p$ to $V p$ reaction, it is better for obtaining qualitative estimates when extracting the absolute value of scattering lengths $\left|\alpha_{V p}\right|$. One finds by comparing the abundant literature that these scattering lengths $\left|\alpha_{V p}\right|$ increase with the meson's mass $m_V$ \cite{Strakovsky:2014wja,Strakovsky:2020uqs,Strakovsky:2019bev,Strakovsky:2021vyk,Wang:2022xpw}. A convincing argument is that the size of the scattering length is related to the radius of the vector meson, which can be written as $r_{q\bar{q}} \simeq 1/ (2 m_{q}) $. As a lighter state compared with the previous research, whether the scattering lengths $\left|\alpha_{\rho^0 p}\right|$ can break the maximum boundary of scattering length $\left|\alpha_{V p}\right|$ is of great interest. Therefore, the present work will allow us to provide more theoretical references for future studies on characterizing the vector meson--proton scattering length by studying the near-threshold photoproduction of vector mesons.

On the experimental side, the SAPHIR Collaboration reported the $\rho^0$ photoproduction cross section at a center of mass (c.m.) energy of up to $E_{\gamma}=$ 2.60 GeV, and the lowest energy of the differential cross section is $E_{\gamma}\in $ [1.4,1.6] GeV \cite{Wu:2005wf}, which is a near-threshold energy. More experimental information on $\rho^0$ photoproduction is essential to gain insight into the internal character of the proton and the $\rho^0 p$ scattering length.
Presently, a JLab experiment is suggested for probing the deepest structure inside the hadron and collecting $\rho^0$ data. High-precision experimental measurements are suggested to be performed at this facility \cite{CLAS:2001zxv}.

This paper is organized as follows. The process of extracting the proton mass radius is provided in Sec.\ref{sec:maa-radius}, and an analysis of the result is presented.
Then, in Sec. \ref{sec:scattering length}, the formalism of scattering length $\left|\alpha_{\rho^0 p}\right| $ from the cross section is introduced, and the results and a discussion are reported. A comparative analysis with the scattering lengths for $\omega p$, $\phi p$, $J/\psi p$, $\psi(2S)p$, and $\Upsilon p$ is suggested. A summary is given in Sec. \ref{sec:summary}.

\section{Scalar GFFs and mass radius}\label{sec:maa-radius}
 Using the standard form, the differential cross section of $\rho^0$ photoproduction
can be written as \cite{Workman:2022ynf,Kharzeev:2021qkd}
\begin{align}\label{eq:differential}
\frac{d\sigma_{\gamma p\to \rho^0 p}}{dt}=\frac{1}{64 \pi W^2} \frac{1}{|{\bf p}_{1 cm}|^2} \left| \mathcal{M}_{\gamma p\to \rho^0 p}(t)\right|^2
\end{align}
 where ${\bf p}_{1 cm}$ is the photon momentum of the $ \gamma p\to \rho^0 p$ process, and $W,t$ is the c.m. energy and momentum transfer, respectively.
Reference \cite{Kharzeev:2021qkd} writes the amplitude $\mathcal{M}$ in terms of the scalar GFFs as
\begin{align}\label{eq:M}
 \mathcal{M}_{\gamma p\to \rho^0 p}=-(Q _u+Q_d) c_2 \frac{16\pi^2 M}{b} \left< P^{\prime}|T|P \right>
\end{align}
where $b=9$ for three light quarks $ u$, $d$, and $s$. $Q_u$ and $Q_d$ represent the coupling of the photon and quarks in the $\rho^0$ meson. We take a superposition of the amplitudes with $Q_u$ and $Q_d$ weighted by the $\rho^0$ wave function and obtain an effective $Q_{quark}=(Q_u-Q_d)/\sqrt{2}=(1/\sqrt{2}) e$.
 $c_2$, the short-distance coefficient, is on the same order of magnitude as $\pi r_q^2$.
The radius of the $q\bar{q}$ pair is $r_{q\bar{q}} \simeq 1/(2 m_{q}) $, with $m_q=m_u=m_d = 0.33$ GeV \cite{Zhao:1998rt}. Finally, the value of $c_2$ is on the order of $\pi r_{q\bar{q}}^2 = 0.28 \text{ fm}^2$. We will check the value of $c_2$ in the subsequent calculation.

\begin{figure}[H]
	\centering
	\includegraphics[width=0.5\textwidth]{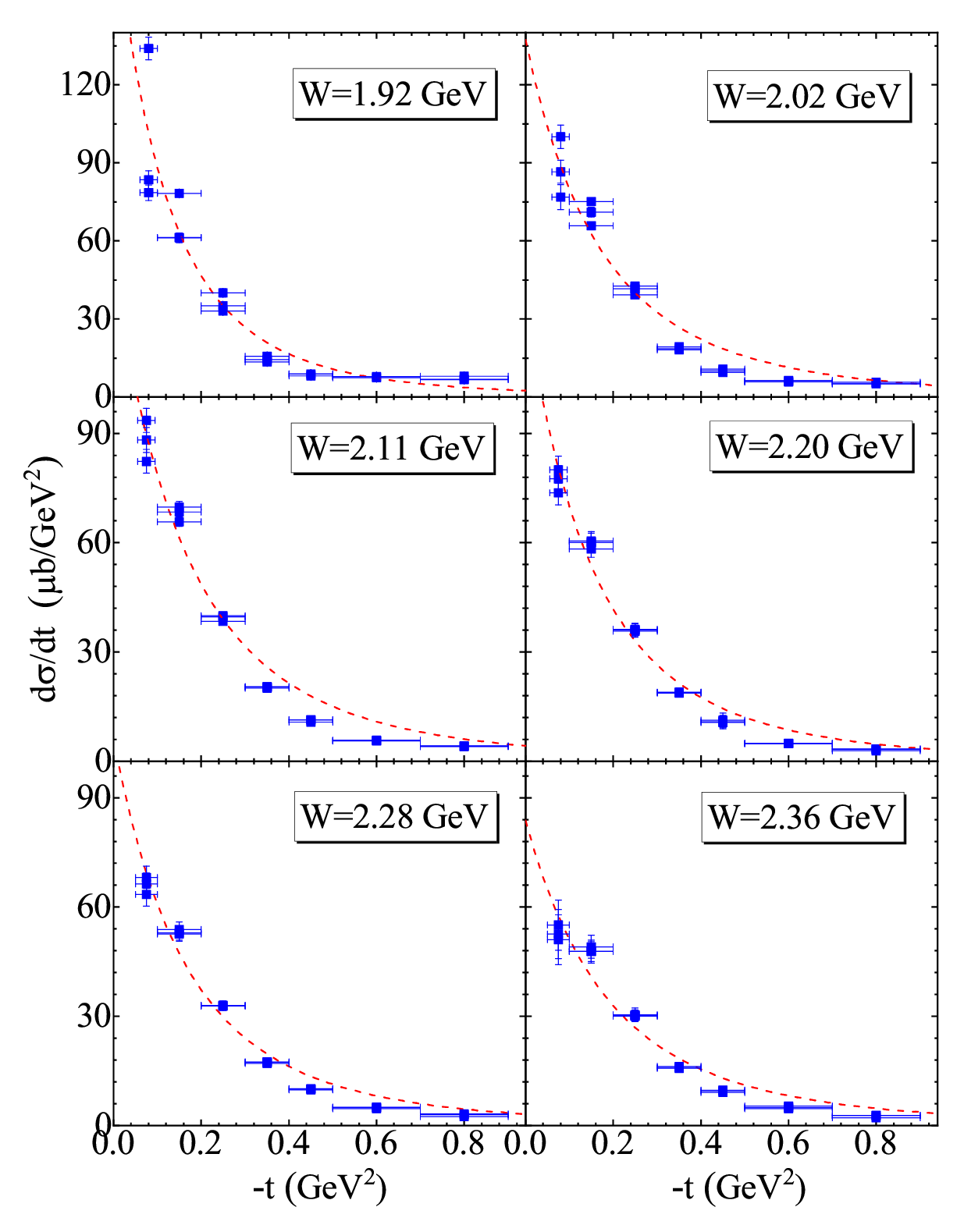}
\caption{Fitted differential cross section as a function of four-momentum transfer squared $ t$ at $W = 1.92$ GeV, $2.02$ GeV, $2.11$ GeV, 2.20 GeV, 2.28 GeV, and $2.36$ GeV, compared to the experimental data \cite{Wu:2005wf}. }
 \label{Fig:other}
\end{figure}

In the weak gravitational field approximation, the proton mass radius is defined through the form factor of the trace of the energy--momentum tensor of QCD \cite{Kharzeev:2021qkd,Wang:2021ujy}.
 The scalar GFFs are usually parameterized in the dipole form and provided as \cite{Kharzeev:2021qkd}
\begin{align}\label{eq:Gtms}
G(t)=\frac{G(0)}{(1-t/m_s^2)^2}
\end{align}
where $G(0)=M$, and $ m_s$ is a parameter adjusted to the experimental data.
Combining
 the scalar GFFs $\left< P^{\prime}|T|P \right>=G(t)$, Eqs. \ref{eq:M} and \ref{eq:Gtms}, the differential cross section of the $\gamma p \to \rho^0 p$ reaction can be determined by fitting the SAPHIR data \cite{Wu:2005wf}.
In this work, the differential cross section at an incident photon energy region $E_{\gamma} = [1.4 \text{ GeV}, 2.6 \text{ GeV}]$ in the current experiment is used to determine the parameters in Eq. \ref{eq:differential}.
 Then, the free parameters $c_2$ and $ m_s$ are fixed by fitting the near-threshold differential cross section. The comparison between the differential cross section and the experimental measurements is manifested in Fig. \ref{Fig:other}, exhibiting a good agreement.

\begin{figure}[H]
	\centering
	\includegraphics[width=0.45\textwidth]{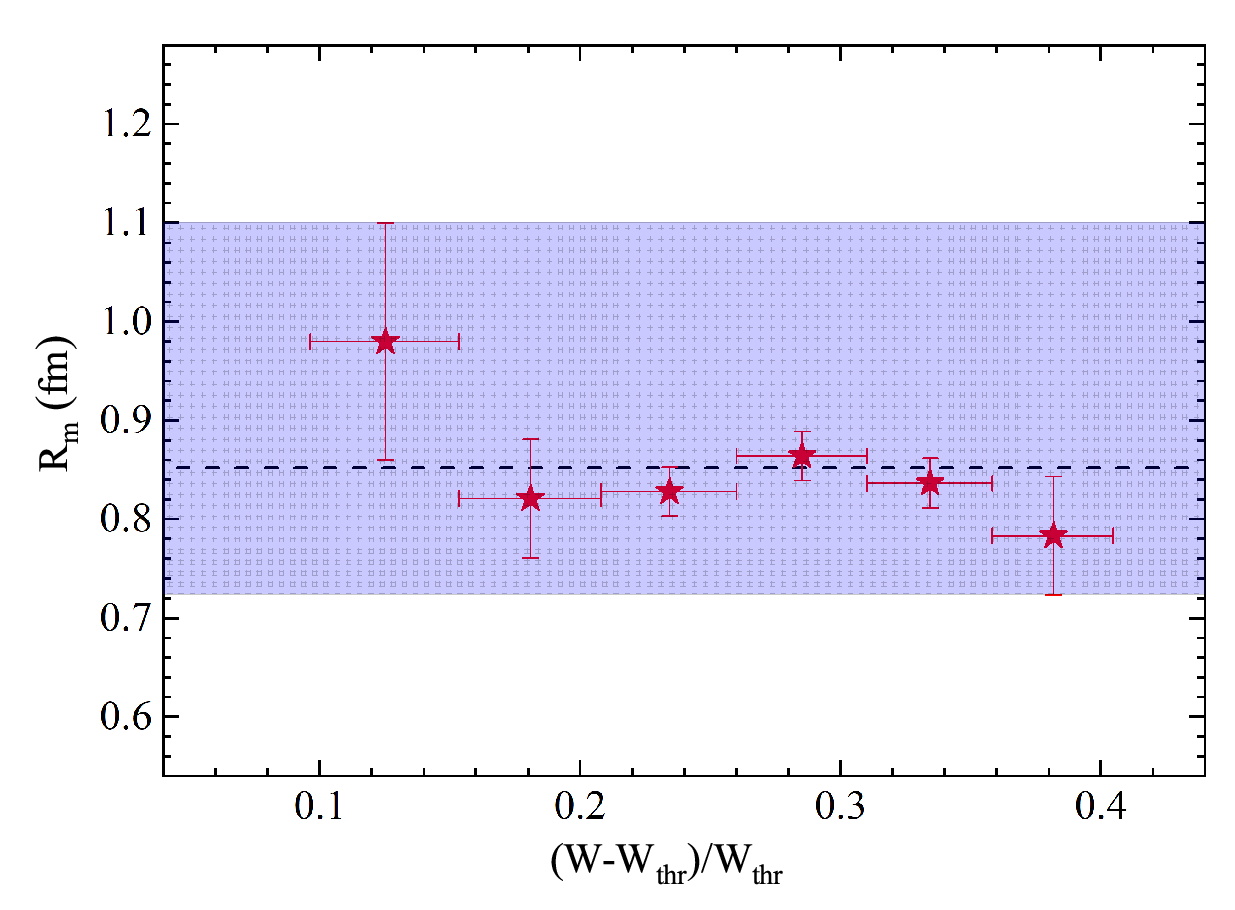}
\caption{Obtained mass radius by fitting the differential cross section of $\rho^0$ photoproduction.
 The red pentagram shows the fitted result at $W = 1.92$ GeV, $2.02$ GeV, $2.11$ GeV, 2.20 GeV, 2.28 GeV, and $2.36$ GeV \cite{Wu:2005wf}. }
 \label{Fig:radius-other}
\end{figure}

 The extracted root-mean-square (rms) short-distance coefficient $c_2= 1.36 \text{ fm}^2$ is on the order of $ \pi r_{q\bar{q}}^2 $, which coincides with our expectation. Note that, compared to the extracted value $c_2=0.207 \text{ fm}^2$ from $J/\psi$ GlueX data \cite{Kharzeev:2021qkd}, a larger short-distance coefficient is extracted from $\rho^0$ photoproduction data.
The parameter $ m_s$ is a more important physical quantity in this work.
We perceive that
the mass radius $R_{m}$ of the nucleon can be defined in terms of the scalar GFFs as \cite{Kharzeev:2021qkd,Miller:2018ybm}
\begin{equation} \label{eq:RG}
\left<R^2_m \right> = \frac{6}{G(0)} \left.\frac{dG(t)}{dt}\right|_{t=0}
=\frac{12}{m_s^2}
\end{equation}
By extracting differential cross section experimental data in c.m. energy $W = 1.92$ GeV, $2.02$ GeV, $2.11$ GeV, 2.20 GeV, 2.28 GeV, and $2.36$ GeV \cite{Wu:2005wf},
 the rms mass radius of the all-fitted result is calculated as $0.85 \pm0.06$ fm, which is shown in the black dashed curve in Fig. \ref{Fig:radius-other}.

 According to Kharzeev's calculation by fitting the GlueX data of $J/\psi$ photoproduction \cite{GlueX:2019mkq}, the mass radius is calculated to be $0.55 \pm 0.03$ fm \cite{Kharzeev:2021qkd}.
Moreover, one work \cite{Wang:2021dis} obtains an average mass radius of $0.67 \pm 0.10 $ fm by extracting the $\phi$ experimental data \cite{LEPS:2005hax} and $0.68\pm0.03 $ fm by extracting the $\omega$ photoproduction data \cite{Barth:2003kv}. Compared with these results obtained by extracting $\omega$, $\phi$, and $J/\psi$ photoproduction, our work for the mass radius is much larger and near the proton charge radius.

\begin{figure}[H]
	\centering
	\includegraphics[width=0.35\textwidth]{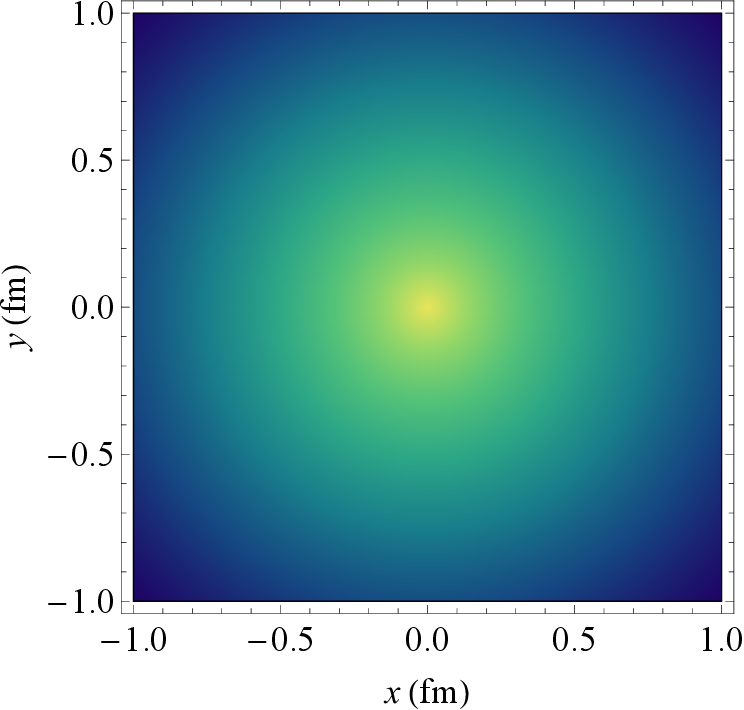}
\caption{2D plot of the radius distribution inside the proton from the Fourier transform of the scalar GFFs $G(t)$.}
 \label{Fig:radius}
\end{figure}

Then, we try to describe the mass distribution inside the proton.
 The Fourier transform of the scalar GFFs $G(t)$ can be written as \cite{Polyakov:2018zvc,Polyakov:2002yz}
\begin{equation}
\begin{aligned}
 \widetilde{D} (r)&=\int \frac{d^3 \Delta }{ (2\pi)^3 } e^{-i\Delta r} G(- \Delta^2)
 \\
&=\frac{ G(0)m_s^3 }{8 \pi } \cdot exp(-m_s r) \\
\end{aligned}
\end{equation}
where $-\Delta^2= t$. Finally, the mass distribution $\widetilde{D} (r)$ inside the proton is derived from $G(t)$, as shown in Fig. \ref{Fig:radius}. Here, the parameters $c_2 $ and $m_s$ in Eqs. \ref{eq:differential}, \ref{eq:M}, and \ref{eq:Gtms} are applied by fitting the $\rho^0$ photoproduction data at $W=1.92$ GeV.
 As shown, the radius distribution is the largest at the center of the proton and then exponentially decreases.

\section{ Scattering length $\left|\alpha_{\rho^0 p}\right|$}
\label{sec:scattering length}

The $\rho^0 p$ scattering length is related to the near-threshold photoproduction of the $\rho^0$ meson. In this paper, the VMD model is used to connect the reactions
$\gamma p \rightarrow \rho^0p$ and $ \rho^0 p \rightarrow \rho^0 p$.
 Applying the effective VMD approach, the near-threshold cross section during the elastic scattering processes becomes \cite{Titov:2007xb,Pentchev:2020kao}
\begin{equation}\label{eq:sigma-total}
\begin{aligned}
 \left. \frac{d \sigma}{dt} \right|_{thr}&= \left.\frac{\alpha\pi^2}{|{\bf p}_{1cm}|^2 g_{\rho^0}^{2} } \cdot \frac{d \sigma^{\rho^0 p \rightarrow \rho^0 p}}{d \Omega}\right|_{thr} \\
&= \frac{\alpha \pi^2}{ |{\bf p}_{1cm}|^2 g_{\rho^0 }^2} \cdot \left|\alpha_{\rho^0 p}\right|^2
\end{aligned}
\end{equation}
Here, the VMD coupling constant $g_{\rho^0}$ is deduced from the leptonic decay width $\Gamma^{\rho^0}_{e^+e^-}$ as \cite{Strakovsky:2019bev}
\begin{align}\label{eq:gV}
g_{\rho^0}=\sqrt{\frac{\pi \alpha^2 	m_{\rho^0}}{3 \Gamma^{\rho^0}_{e^+e^-} }}.
\end{align}
Here, $\Gamma^{\rho^0}_{e^+e^-} = 7.04$ keV from the Particle Data Group \cite{Workman:2022ynf}, so $g_{\rho^0} = 2.47$ is obtained for the $ \rho^0 $ meson.
By setting $ \left. \frac{d \sigma}{dt} \right|_{thr}=b_1$,
the absolute value of scattering length $\left|\alpha_{ \rho^0 p}\right|$ is given as
 \begin{align}
\left|\alpha_{\rho^0 p}\right|= \frac{g_{\rho^0} |{\bf p}_{1cm}| }{\pi} \sqrt{\frac{b_1}{\alpha}}.
\end{align}
Note that $b_1$ and $|{\bf p}_{1cm}| $ must satisfy the conditions that are obtained from the threshold energy.

\begin{figure}[H]
	\centering
	\includegraphics[width=0.45\textwidth]{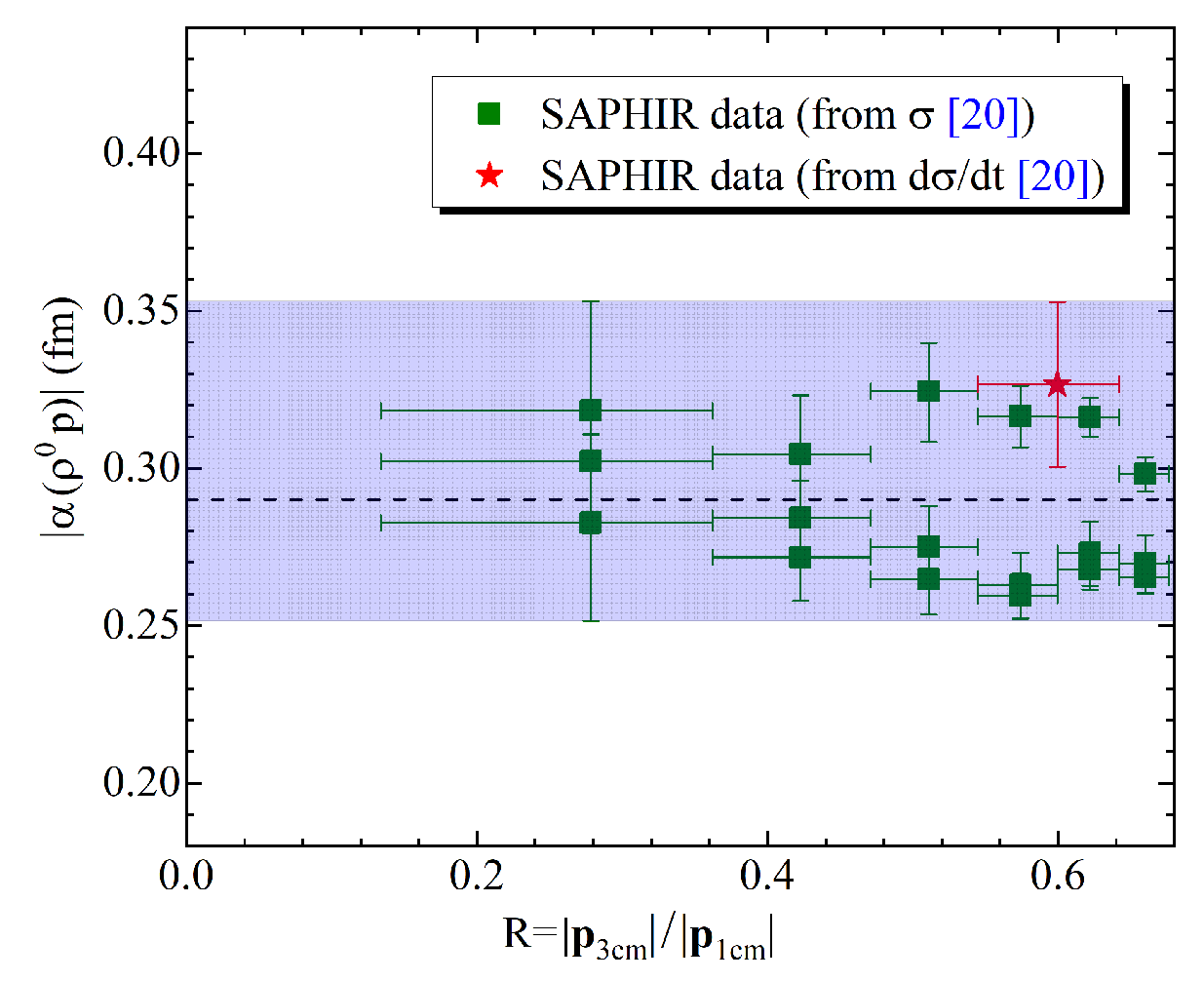}
\caption{Obtained absolute value of the scattering lengths of the $\rho^0$--proton interaction.
 Here, the red pentagrams and green squares are extracted from SAPHIR differential and total cross section data, respectively \cite{Wu:2005wf}. The light-blue band reflects an error bar of $\left|\alpha_{\rho^0 p}\right|$.}
 \label{fig:scattering length}
\end{figure}

According to the fitted results in Fig. \ref{Fig:other},
the differential cross section $ \left. \frac{d \sigma}{dt} \right|_{thr} $ of $\rho^0$ photoproduction can be obtained. Thus, the absolute value of the rms $\rho^0 p$ scattering length can be estimated as $\left|\alpha_{\rho^0 p}\right|= 0.32 \pm0.03 $ fm, as shown in Fig. \ref{fig:scattering length}.

The scattering length $\left|\alpha_{\rho^0 p}\right|$ can also be expressed by the total photoproduction cross section \cite{Wang:2022xpw}
\begin{equation}\label{eq:sigma}
 \left. \sigma \right|_{thr}\left(R\right) = R \cdot \frac{4 \alpha\pi^2}{g_{\rho^0}^{2}} \cdot\left|\alpha_{\rho^0 p}\right|^{2}
\end{equation}
where $R$ is the ratio between the final momentum $|{\bf p}_{3cm}|$ and the initial c.m. momentum $|{\bf p}_{1cm} |$.
As a function of c.m. energy, $ R(W)$ has a range of $R(W)\in[0,1)$ and is positively correlated with the c.m. energy. Thus, the absolute value of
 scattering length $\left|\alpha_{\rho^0 p}\right|$ is given as
\begin{align}\label{eq:alpha}
\left|\alpha_{\rho^0 p}\right| = \frac{g_{\rho^0 }}{2\pi} \sqrt{\frac{ \sigma(R) }{\alpha R}} .
\end{align}
 We extract the scattering length from the total cross section SAPHIR data at an energy region $W\in$[1.72 GeV,~2.00 GeV] using the S$\ddot{\text{o}}$ding, Ross--Stodolsky, and Berit--Winger methods \cite{Wu:2005wf}, as shown in the green squares in Fig. \ref{fig:scattering length}.
 The rms absolute value of the scattering length of the proton and $\rho^0$ meson is calculated as $0.29\pm 0.02$ fm, which is consistent with the result extracted from the differential cross section if the error bar is considered. Finally, combined with the differential and total cross sections, the absolute value of the $\rho^0$--proton scattering length $|\alpha_{\rho^0 p}|= 0.31 \pm 0.06 \text{ fm}$ is obtained considering the error bars of the all-fitted result.

\begin{figure}[H]
	\centering
	\includegraphics[width=0.45\textwidth]{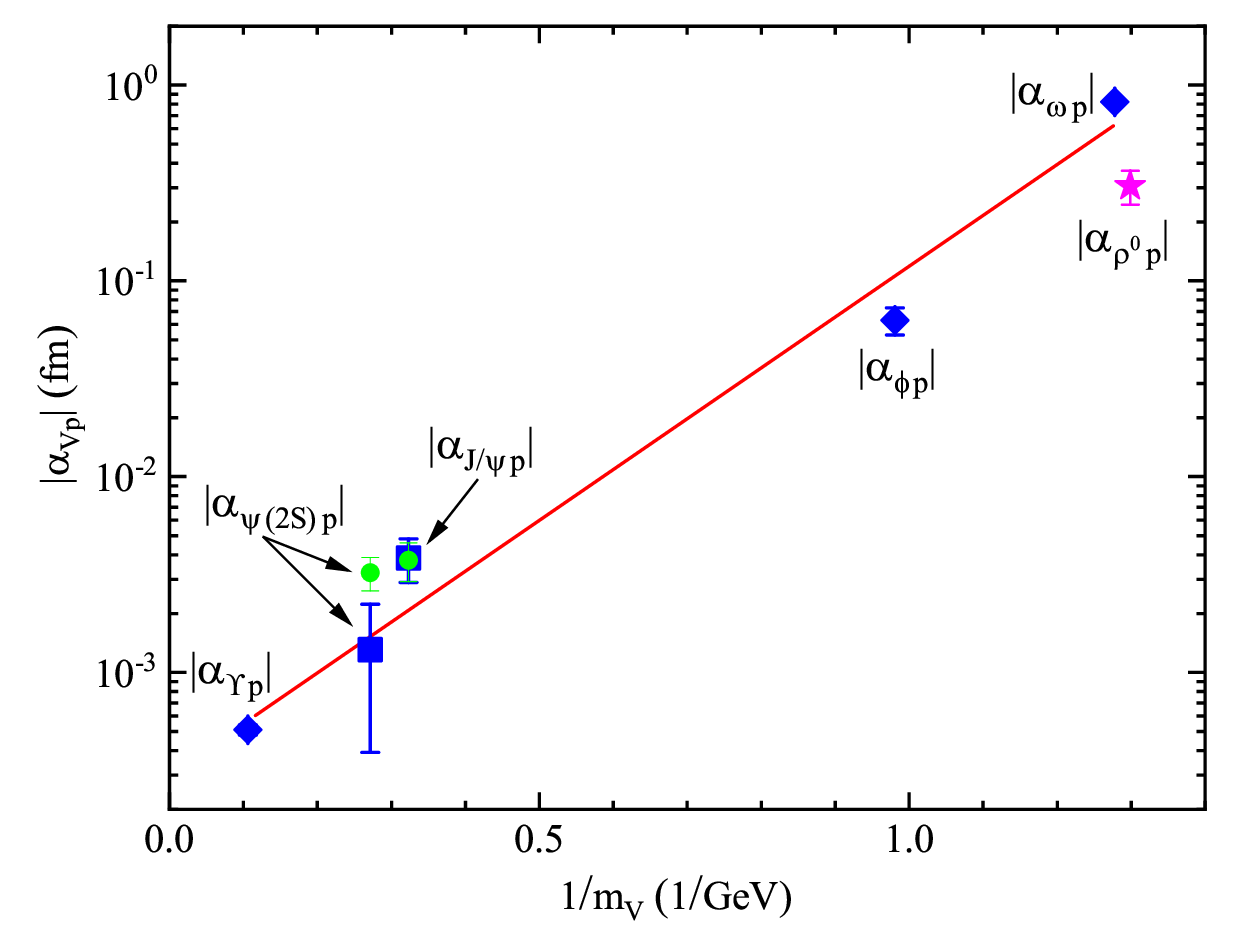}
\caption{Comparison of the absolute value of scattering lengths $\left|\alpha_{V p}\right|$ as a function of the inverse mass of vector mesons, including $\rho^0, \omega, \phi, J/\psi,\psi(2S),$ and $ \Upsilon$. The magenta pentagram shows the absolute value of the $\rho^0 p$ scattering length from this work.
The blue rhombus shows the analysis results of $ \omega, \phi,$ and $ \Upsilon$ \cite{Strakovsky:2014wja,Strakovsky:2020uqs,Strakovsky:2021vyk}. The blue squares and green circles show the absolute value of $J/\psi p$ and $ \psi(2S) p$ scattering lengths from the two-gluon exchange model and the effective Pomeron model, respectively \cite{Wang:2022xpw}.
The solid red line is hypothetical \cite{Strakovsky:2021vyk}. }
 \label{fig:alpha-V}
\end{figure}

On the basis of the recent threshold measurements of the photoproduction of $ \omega$ and $ \phi $ mesons off the proton by the A2 (MAMI) \cite{Strakovsky:2014wja} and CLAS (JLab) \cite{Dey:2014tfa}, one can determine the absolute value of vector meson--proton scattering lengths $\left|\alpha_{\omega p}\right|$ and $\left|\alpha_{\phi p}\right|$ using the VMD model \cite{Strakovsky:2014wja,Strakovsky:2020uqs}. Moreover, the absolute value of the $\Upsilon p$ scattering length is studied using quasi data generated from the QCD model \cite{Guo:2021ibg,Strakovsky:2021vyk}. In our previous work \cite{Wang:2022xpw}, the $J/\psi p$ and $ \psi(2S) p$ scattering lengths are studied systematically. Finally, the relationships of $ \omega, \phi, J/\psi, \psi(2S), ~\text{and}~ \Upsilon$ can be determined as
\begin{align}\label{eq:alpha}
\left|\alpha_{\Upsilon p}\right| < \left|\alpha_{\psi(2S) p}\right| < \left|\alpha_{J/\psi p}\right| < \left|\alpha_{\phi p}\right| < \left|\alpha_{\omega p}\right| .
\end{align}
Note that the absolute value of the $\rho^0$--proton scattering length $\left|\alpha_{\rho^0 p}\right|$ disobeys the rule of Eq. \ref{eq:alpha}, as shown in Fig. \ref{fig:alpha-V}.
 This deviation is mainly due to the $\rho^0$ meson being regarded as a point in the analysis. This result can simply be considered a first- or second-order approximation ignoring the broad width of the $\rho^0$ meson.

Using the QCD sum rule, one work \cite{Koike:1996ga} obtains the
absolute value of the $\rho^0 p$ scattering length as $0.47\pm 0.05$ fm. In fact, if the error is considered, our results are close to the prediction of the QCD sum rule.

\section{Summary}\label{sec:summary}

In this paper, the proton mass radius and scattering length $\left|\alpha_{\rho^0 p}\right| $ are extracted for the first time by fitting the experimental data of rho-meson photoproduction. The value of the rms mass radius is calculated as $0.85\pm 0.06 \text{ fm } $ from the differential cross section at the near-threshold c.m. energy. One finds that the proton mass radius extracted from $\rho^0$-meson photoproduction is larger than the other results extracted by $\omega$, $\phi$, and $J/\psi$ photoproduction but is near the proton charge radius. Unlike the heavy quarkonium, for the $\rho^0$ meson, one may no longer assume the dominance of the scalar gluon operator in the production amplitude, as the scalar quark--antiquark pair in the $t$-channel should give a large contribution. From this viewpoint, our extracted result probably represents the quark radius of the proton, which may explain why the extracted radius is very near the charge radius dominated by quarks. Combined with the results from light meson $\phi$ photoproduction \cite{Wang:2022uch}, one finds that the estimated mass radius is also similar to the proton charge radius.

Further, under the VMD model framework, the absolute value of the $\rho^0$--proton scattering length is calculated to be $|\alpha_{\rho^0 p}|= 0.31 \pm 0.06 \text{ fm}$. Apparently, the scattering length $\left|\alpha_{\rho^0 p}\right|$ disobeys the rule that the absolute value of the vector meson and proton scattering length $|\alpha_{V p}|$ increases with the meson's mass. The deviations can be attributed to the point-like object assumption of the $\rho^0$ meson
 and the neglect of the broad width of the $\rho^0$ meson in the analysis.

Our results should provide an important numerical reference for subsequent research in this area. Moreover, because of the large error in the SAPHIR data we used, more high-precision experimental measurement data is very much needed, which can be realized in the JLab experiment \cite{CLAS:2001zxv}.

\Acknowledgements{We greatly appreciate the valuable discussion with Prof. Dmitri E. Kharzeev on proton mass radius. We are also very grateful for the effective discussion with Prof. Igor I. Strakovsky on scattering length. This work is supported by the National Natural Science Foundation of China under Grant Nos. 12065014 and 12047501, and by the Natural Science Foundation of Gansu province under Grant No. 22JR5RA266. We acknowledge the West Light Foundation of The Chinese Academy of Sciences, Grant No. 21JR7RA201.}


\end{multicols}
\end{document}